# Probably Reasonable Search in eDiscovery

Herbert L. Roitblat, Ph.D.

## Abstract

In eDiscovery, a party to a lawsuit or similar action must search through available information to identify those documents and files that are relevant to the suit. Search efforts tend to identify less than 100% of the relevant documents and courts are frequently asked to adjudicate whether the search effort has been reasonable, or whether additional effort to find more of the relevant documents is justified. This article provides a method for estimating the probability that significant additional information will be found from extended effort. Modeling and two data sets indicate that the probability that facts/topics exist among the so-far unidentified documents that have not been observed in the identified documents is low for even moderate levels of Recall.

Although the Federal Rules of Civil procedure are quite clear that eDiscovery does not require a perfect search, but only a reasonable one, it is often a point of contention as to just what constitutes a reasonable search. One reason for this contention is the challenge of predicting the value of additional search effort.

eDiscovery professionals have grown accustomed to measuring search accuracy, particularly with the introduction of computer assisted review. When computer-assisted (machine learning) review was first introduced to the legal community around 2010, the courts and the attorneys sought assurance that the computer was doing an effective job. It was easy to show then, that the computer systems were at least as accurate as human reviewers. (Roitblat, Kershaw, & Oot, 2010). Outcome measurement was introduced in this context and has continued to be a significant factor since then.

In fact, computer-assisted review (also called technology-assisted review, TAR, or predictive coding) has consistently been measured to be more accurate than human review, and more accurate than simple keyword search. These measurements, however, usually show that the search process is still less than perfect. Recall, the most common measure of the completeness of the discovery process, is typically less than 100%, meaning that some relevant documents have been missed. Incompleteness leads some to fear that critical information might also have been missed. This fear of missing out is supported by anecdotes of subsequently finding additional important information when the remaining documents are later examined. As statisticians are fond of saying, however, anecdotes are not data.

In cases with less than perfect Recall, judges face the quandary that they know that relevant documents are missing following a certain amount of effort. There is a possibility that extending the search effort might provide additional information, but there is uncertainty about whether that effort can be justified by the value of the additional information that might be found. In this paper we examine systematically the likelihood of there being significant additional information remaining after a search.

There are three reasons why novel information might be found with extended effort. (1) New information requirements could be uncovered, during the review process. A party might find that there are new questions to be answered. (2) Documents that that have been seen might contain important information, but that importance may not have been recognized. For example, many documents may mention the word "chocolates," but only after substantial review do the attorneys recognize that "chocolates" was code for "bribe" (US Securities and Exchange Commission, 2012) (3) Critical information might not be encountered because the search process has been imperfect. This paper only concerns the third situation. What is the likelihood of novel relevant information in the documents that remain after a putatively reasonable but imperfect search?

## Why this analysis is important to eDiscovery

Many of the ideas about reasonableness in eDiscovery depend on proportionality. In general, the court considers the burden of an effort, such as continuing to review documents against the value/importance that can be expected from that effort. In this context, the main question is the amount of new information that could be expected to be gained compared with the cost of getting it. Searching for and reviewing documents takes time, effort, and money.

In the absence of systematic evaluation, the courts and the parties are left to their intuition to predict the probability and value that can be expected from effort. People, in general, are poor at reasoning about probabilities and thus may come to inappropriate conclusions.

Which of these is most likely to be true? (a) The next person you see will be a bald man, or (b) The next person you see will be a man. Logically, it must be true that encountering a man is more likely than encountering a kind of man (bald man), but people often rate statements like the first as being more probably true than the second (Kahneman & Tversky, 1983).

The so-called birthday paradox is another example of poor intuitive probability estimation (Flajolet, Gardy, & Thimonier, 1992). How many people would have to attend a party to find at least two people that shared the same birthday? How many people would have to attend the party to find at least one person born on each day of the year? Incorrect intuitions about this second question are very important to eDiscovery, as you will see. The answers will be given later.

For example, people over-estimate the likelihood of things that were heard about recently, or things that are easier to remember. This tendency is called the availability bias. People over-estimate the likelihood of information that is easier to pull from memory (Kahneman, 2011).

If the judge's intuition overestimates the probability and value of prospective additional information, then the court might be persuaded to require an unreasonable amount of work for meager returns. A simple analysis provides a principled prediction of that likelihood of finding additional previously unknown information. This systematic analysis can be used to calibrate expectations. It can counteract the inherent fear of missing out (FOMO) on this potential additional information.

In one of the first studies of the effectiveness of using computerized search in eDiscovery, Blair and Maron (1985) worked with attorneys on keyword searching. "The information-request and query formulation procedures were considered complete only when the lawyer stated in writing that he or she was satisfied with the search results for that particular query (i.e., in his or her judgment, more than 75 percent of the 'vital,' 'satisfactory,' and 'marginally relevant' documents had been retrieved)" (p. 291). Interestingly, when they eventually evaluated the success of these searches, Blair and Maron found that the lawyers' queries had actually resulted in 20%, not 75%, Recall.

The Blair and Maron study is often cited as an example of the poor accuracy of keyword searching, but it is also an indicator that lower levels of Recall may be sufficient to provide all of the necessary information in a case. The responsible attorneys were satisfied that 20% Recall met their information need, even though it was missing many relevant documents.

Roitblat, Kershaw, and Oot (2010) compared two human review teams and two machine learning systems intended to identify responsive documents. The two human teams achieved 49% and 54% Recall respectively. The analysis described in this paper (the FOMO analysis) seeks to formalize this general acceptance of imperfect search. Lawyers were satisfied with search, at least in part, because the searches they conducted did, in fact, identify all of the relevant information, even if they failed to identify all of the relevant documents.

## The FOMO analysis

The key to the FOMO analysis is the recognition that the information contained in a document can be considered separately from the document itself. A search may identify less than all of the relevant documents, but at the same time identify all of the relevant information contained in those documents. The same information may be contained in multiple documents. Each document may contain one or more topics and the same topics may appear in one or more documents.

A putatively reasonable search may leave documents undiscovered without leaving topics or facts undiscovered. The analysis says nothing about the content of those topics. They could be as general as the kind of issue tags that one might apply during an eDiscovery review, or they could be as specific as individual facts (e.g., Promeca employee X paid $Z to Instituto Mexicano del Seguro Social, employee Y).

Facts or topics can vary in their probative value and in their prevalence. The analysis concerns the probability of each fact or topic, not its content or importance. As I describe more completely later in this paper, parties often do not know what the specific topics are or even their probability, but they can still use this approach to understand what they are likely to gain from further search effort.

The math of the FOMO analysis is quite simple and has a very long history. Put simply, the main statistical question is: What is the probability of not finding an item in a big set and then finding it in a smaller set? The big set is the set of relevant documents that the search method, for example, a keyword search or a machine learning method, has identified. The smaller set is the set of relevant documents that were missed by the search process. If we want to estimate the value of a continued search effort, how likely are we to find some piece of information in the missed set that we have not seen already in the identified set?

## A mini tutorial on probability using dice

Craps is a dice game in which a "shooter" rolls a pair of dice and records the sum of the numbers that show on top of each die at the end of the roll. The total can be any number between 2 (both dice show 1) and 12 (both dice show 6). Some numbers are more commonly rolled than others. There is only one way to end up with a sum of 2 and only one way to end up with a sum of 12, on the other hand, the shooter can hit a 6 in five ways (1 & 5, 2 & 4, 3 & 3, 4 & 2, 5 & 1).

For our purposes, let's consider a simplified game where the question is how many rolls does it take, on average, to hit all of the numbers from 2 to 12 (61.22 rolls on average). This general kind of dice problem has been studied since the 17$^{th}$ Century.

Each roll of the dice is analogous to a document and the sum of the two dice is analogous to the information it contains. The same sum appears in multiple rolls. When all 11 sums have been recorded, there are none left to be found. The general form of this problem is called the Coupon Collector's problem. How many packs of bubblegum do you have to open to find one example of each player's trading card (Flajolet, Gardy, Thimonier, 1992)? Or how many boxes of cereal do you need to buy to find a complete set of the coupons where each box contains one coupon. How many documents do you need to examine to find at least one example of every topic? The coupon collector's problem was first described by A. De Moivre in 1708 (see Ferrante & Saltalamacchia, 2014).

The first roll of the dice is guaranteed to provide a number that has not been seen before, because at that point, no numbers have yet been seen. The shooter rolls the dice and adds the sum to her list of seen numbers. The second roll could match the first one, but it is more likely to be different. The third could match either the first, or the second (if they are different) or add another number to our list of seen numbers, and so on. In the case of dice, and in the case of eDiscovery generally, the coupons may differ in probability. Some coupons may be more common than others, this difference complicates the computations, but the principle is the same.

In eDiscovery, the number or probability of the topics/coupons/facts may not be known. Instead, we may know the number of relevant documents that have been identified and the Recall level for the search that identified them. We may also have some idea about the relative frequencies of potential topics. I will present some empirical evidence about this distribution later. Like many things (e.g., [words, city sizes, family names](words, city sizes, family names)), the topics are likely to follow a pattern called Zipf's Law (Zipf, 1935; Clauset, Shalizi & Newman, 2009 ).

For example, an examination of Shakespeare's plays shows that most of his writing consists of a small number of words. About 79% of the words used in his plays consist of the same set of 998 (4%) unique words. Conversely, most of his vocabulary was used rarely. A total of 17,089, (69%) of the unique words, each appeared 10 or fewer times, accounting collectively for only 7% of the total words used (Culpeper, 2007, 2011). A small number of unique words constituted a large proportion of the total words and a large number of unique words accounted for a small proportion of the total words. Topic occurrence, like word occurrence, is very unevenly distributed.

When the topics differ substantially in probability the coupon collector's problem becomes much more complex (Zoroa, Lesigne, Fernández-Sáez, Zoroa & Casas, 2017), but it is dominated by the prevalence of the rare topics. The time needed to collect all of the coupons/topics will depend strongly on the time needed to find the rarest of them. In the next section, we will discuss some actual collections and their topics.

## Observations of topics and collections
### Study 1
This study relied on a relatively small set of several thousand messages that had been carefully analyzed for the presence of racial micro-aggressions. More details on this study are available in Roitblat (2020). It used unsupervised machine learning (Latent Dirichlet Analysis; Blei, Ng, & Jordan, 2003) to identify 100 topics and one of several realistic independent machine learning methods to assign messages to the identified and to the missed sets. The two most important methods were Bayesian categorization and Continuous Active Learning (Cormack & Grossman, 2014). This later method presents the documents for review and decision-making in a non-random order (as confirmed by a statistical analysis).

The documents for this study consist of a set of micro-aggression communications collected by Breitfeller et al. (2019), which they carefully classified for relevance (Microaggression or not). Microaggressions are statements or actions, that express a prejudiced attitude toward a member of a marginalized minority.

A few example microaggressive statements:

- At least I don't sit on my ass all day collecting welfare, I EARN my money.
- Did you get this job because you're pregnant?
- Do your parents make sacrifices so that you can go to our school?

Both machine learning classifiers were trained without using any of the topic categories to recognize these microagressions and each achieved approximately 80% Recall. There was no indication that the non-random order in which document decisions were made by the Continuous Active Learning method made a significant difference in the probability that a given topic would be identified or missed because each topic was represented in the identified set. All of the topics identified by the Latent Dirichlet Analysis were present within the identified set, so no new topics were available to be found among the 20% of messages that were misclassified as not relevant. Documents were missed, but not topics. In this set, the most common topic appeared in 1.5% of documents. The least common topic appeared in 0.1% of the documents.

Study 2

I also examined a relatively large set of documents and their topics. This set involved just under 12 million web pages categorized into 64 different topics by a commercial system. Each of these documents could be considered "relevant." Each page could be classified into more than one topic, averaging about 1.37 topics per page. The topics varied widely in the number of documents that contained them, and thus in their probability. The most common topic appeared in about 36% of the documents (about 1 in 3). The rarest topic occurred in about 0.00014% (about 1 in 714,286 pages).

The probability of a topic, meaning its relative prevalence, is the key factor that determines how many documents will have to be examined to find it. The math of this process is actually very simple (Feller, 1968). Discovering the topic in nth document, depends on non finding it in any previous document. The first examined document either contains the topic or it does not. If the first document contains the topic, then that topic is no longer novel. If it does not contain the topic, then the second one might, and that topic is exhausted after 2 documents, and so on.

The probability of finding it in the second document is the probability of not finding it in the first one and then of finding it in the second one. The probability of finding the topic in the 100$^{th}$ document depends on not finding it in the first, the second, the third and so on up to the ninety-ninth, and then finding it in the 100$^{th}$.

For the most common topic, the probability of finding it in the first document is 36%, so we could expect that on 36% of our searches, we would be done with this topic after one document. The probability of not finding it in the first documents is 64% (100% - 36%), so the probability of finding it in the second document is the probability of not finding it in the first and the probability of then finding it in the second (64% * 36% = 23%). The probability of finding in the third follows the same logic (64% * 64% * 36% = 15%). That's all there is to it. The rest is bookkeeping.

Figure 1 shows the number of categories that were observed as additional pages were scanned in the "accession" or original order in which the pages were encountered in the ordinary course of business. Most of the topics were observed during the first few documents. All of them were observed within 931,226 documents. As expected, the final topic to be observed was the usually lowest frequency one.

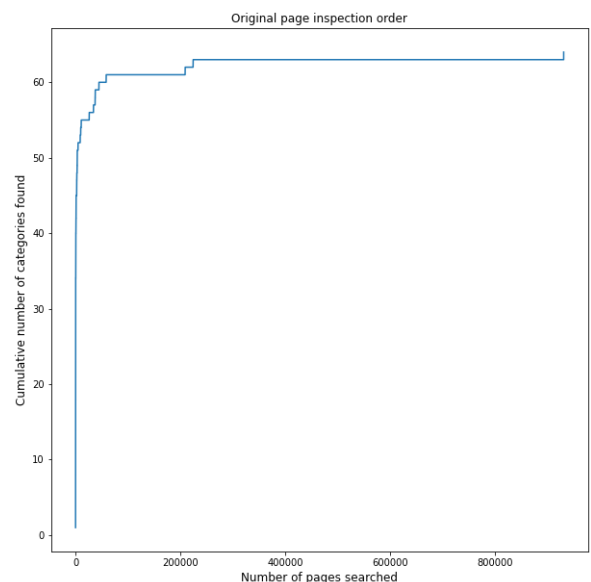

*Figure 1. The cumulative number of categories observed as documents were scanned.*

In the ordinary course of business, the rarest topic was encountered after only about 8% of the documents had been categorized, corresponding to 8% Recall. That is, even though only 8% of the relevant documents had been "discovered" so far, 100% of the topics had been encountered.

But one series does not allow us to draw any general inferences concerning the probable number of documents that would have to be searched. For that, I ran some simulations, which consisted of randomly shuffling the pages and counting the number of documents that would have to be reviewed to find all 64 topics. Each shuffle would correspond to a single eDiscovery case, because any individual case would have only one document ordering (as in the ordinary course of business). Any one case provides one shuffle of the documents, so each repeat shuffle would correspond to a new eDiscovery case. All of the topics were found in every shuffle.

I repeated this shuffling and search process 2,000 times. Figure 2 shows the number of shuffles in which the corresponding number of documents were searched to find all 64 topics. The first bar shows that on 213 of the shuffles, all 64 topics were found after evaluating between 56,381 and 309,789 documents. The second bar shows that on 435 of the shuffles, the complete set was found after 309,789 to 563,198 pages. The final bar shows that one shuffle took more than 4,871,144 and less than 5,124,553. The maximum number of documents needed to find at least one example of each of the 64 topics was 5,124,553 over all 2,000 shuffles.

In 10% of the shuffles, it took 301,028 documents or fewer to find at least one example of each topic. In 20% of the shuffles, it took 413,766 shuffles or less to find them all. In half the shuffles (the 50$^{th}$ percentile), the complete set was found in 797,179 documents or less. In 95% of the shuffles, the complete set was found in 2,202,935 or fewer scanned documents.

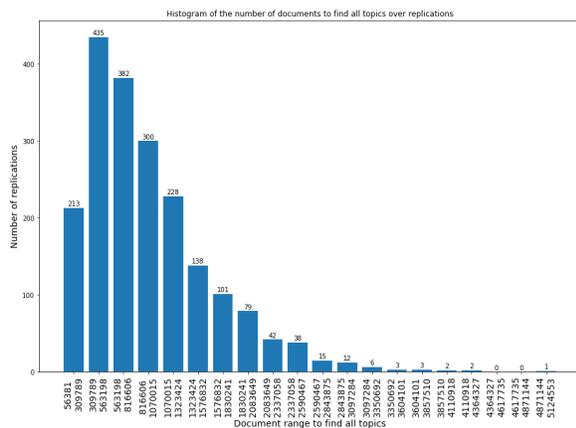

*Figure 2. Histogram of the number of documents that needed to be searched to find all 64 topics after a random shuffle.*

The 2,202,935 documents needed to satisfy 95% of the searches corresponds to 18.4% Recall (2,202,935/11,971,064 = 18.4%). Thus, if about 2 million documents, were reviewed, it is 95% likely that all of the relevant topics would have been identified, given that the lowest probability topic was no lower than 1 in 714,286.

### Back to eDiscovery

Mathematical analysis, backed by observations with two document collections consistently points to the idea that there is a low probability that significant unique information will be missed if there is a reasonably-sized eDiscovery production. It does not seem to matter substantially whether the documents were analyzed in random order or in order of repeated predictions relevance, the final set of documents appears to be what is critical to finding the complete set of topics. In active learning, the ranking of the documents is changed after each batch and the resulting order is not very much different from random (as measured by entropy).

In deriving and testing the FOMO models, we needed and had access to information that is not available in eDiscovery. We could not evaluate the quality of a prediction without having some baseline level of truth against which to compare the prediction. In the two data sets, we knew what the topics were and the probability that a selected document would have that topic. In eDiscovery, we do not generally know the content of the topics, their number, or their probabilities. We do not know whether a given topic is even in the collection. But we can extrapolate from the situation where we know what the truth is to assert that the same proportions would apply whether we know the "truth" ahead of time or not.

Fortunately, the analysis does not concern itself with the content of the topics, and we can rearrange the modeling equations to take what we do know and derive predictions from that. One of the things that we typically do know in eDiscovery is the Recall level and, as we will see, that comes in handy. As a result, we can ask, what is the probability that a topic was missed in the identified set but will be found in the missed set?

As noted, the lower the probability of a topic, the more documents must be examined on average to find it. Conversely, the more documents that have been examined, the lower the probability of any topic that might have been missed. If topic is not found in the first document and not found in

Table 1, below, I have estimated the confidence of there being a novel topic (that is, one that was not found in the identified set of documents) for a specific number of eDiscovery scenarios involving three different sizes of production and four different levels of Recall. As we would expect, Recall is the main determinant of confidence in having conducted a reasonable discovery process.

The results in this table assume a confidence level of 95% in estimating the probability of missing a topic. A confidence level of 95% means that if we repeated an estimation process a hundred times, 95 of those times we would find an estimate within the range of the confidence interval. This confidence interval ranges from the second, and so on, we can estimate its probability after finding it in a hundred, a thousand, or a hundred thousand documents. We can say with 95% confidence, for example, that if we had 2.2 million documents in our production (that were relevant), then any missing topic would have to have a probability less than 1 in 714,286 of the relevant documents.

In

0.0 up to the estimated prevalence of the topic. In this context, a confidence level of 95% means that in 95% of eDiscovery cases with the corresponding number of documents and the corresponding level of Recall, a novel topic should appear in the missed set (the subset of relevant documents that were not identified) with the probability in the last column. The higher the Recall level, the lower is this probability, but even at 50% Recall we can expect to find a novel topic in the missed set in less than 5% of the cases. At 50% Recall, we can be 95% confident that we will not miss any significant unique information. We will return to this claim in a moment.

Table 1. The confidence of there being a novel topic in the missed set as a function of Recall and the size of the identified set.

| Number of documents produced | Confidence | Estimated prevalence of topic | Recall | Number of relevant documents missed | Prob. topic in missed set | Confidence of there being a novel topic in missed set |
|---|---|---|---|---|---|---|
| 50000 | 0.95 | 0.0060% | 80% | 12500 | 52.71% | 2.636% |
| 100000 | 0.95 | 0.0030% | 80% | 25000 | 52.71% | 2.636% |
| 200000 | 0.95 | 0.0015% | 80% | 50000 | 52.71% | 2.636% |
| 50000 | 0.95 | 0.0060% | 70% | 21428 | 72.30% | 3.615% |
| 100000 | 0.95 | 0.0030% | 70% | 42857 | 72.30% | 3.615% |
| 200000 | 0.95 | 0.0015% | 70% | 85714 | 72.30% | 3.615% |
| 50000 | 0.95 | 0.0060% | 60% | 33333 | 86.43% | 4.321% |
| 100000 | 0.95 | 0.0030% | 60% | 66666 | 86.43% | 4.321% |
| 200000 | 0.95 | 0.0015% | 60% | 133333 | 86.43% | 4.321% |
| 50000 | 0.95 | 0.0060% | 50% | 50000 | 95.00% | 4.750% |
| 100000 | 0.95 | 0.0030% | 50% | 100000 | 95.00% | 4.750% |
| 200000 | 0.95 | 0.0015% | 50% | 200000 | 95.00% | 4.750% |

The reader may be wondering why the number of produced documents does not have an effect. The answer is that for each level of Recall, the number of documents in the missed set is always the same proportion of the number of documents in the identified set. At 80% Recall, 80% of the relevant documents are contained in the identified set and 20% of the relevant documents are contained in the missed set. At 50% Recall, the identified set

Table 1 refer to **relevant** documents. In the identified set, these are the documents that have been identified as

Table 1 refers to there being a new topic among these missed **relevant** documents, but it does not speak to how those relevant documents are to be identified. They must first be identified and separated from the non-relevant documents that usually make up the bulk of a collection. Given that the Table 1 is not the proportion of total documents that contain novel information. It is not the estimated prevalence of the topic in the entire set. It is the probability of finding at least one example of a document with a topic if all of the relevant documents in the missed set were to be identified. Identifying a single novel topic among the missed set may require a prodigious effort to first locate and recognize the relevant documents and then further analyze these to identify a novel topic among the mostly familiar ones.

## Conclusion

Exactly how much effort is justified in a specific case will depend on the characteristics of that case and on costs and benefits associated with that effort. We can fail to make a proper assessment of that tradeoff if we under-estimate the effort needed or if we over-estimate the likely benefit from that effort. The data of this paper and the FOMO model cannot directly answer the question of whether more effort is justified, but they can help to inform reasonable decision making about that effort. In productions containing more than 50,000 documents, the expected prevalence of a novel topic is less than 1%. Even at 50% Recall, there is less than a 5% chance that even one more piece of significant unique information will be found through extended effort.

and the missed sets both have the same number of documents.

The probability of missing a topic decreases as the size of the production set increases, but correspondingly, the probability of finding in the missed set also decreases because the size of the missed set also declines.

*The document counts in*

relevant. In the missed set, however, the identify of these documents is unknown. The final column in documents containing a purported novel topic have not yet been found, they may be extremely difficult to identify among the many remaining documents. Also, keep in mind that the last column in

Finally, the birthday problem. It takes about 24 people to attend a party on average, for at least two of them to share a birthday. It takes about 2,365 people on average to collect at least one person with a birthday on each day of the year. Both estimates make the simplifying assumption that births are equally distributed throughout a 365-day year.